# Coexistence of polar distortion and metallicity in PbTi$_{1-x}$Nb$_x$O$_3$

Jun-xing Gu[1], Kui-juan Jin[1,2]*, Chao Ma[1], Qing-hua Zhang[1], Lin Gu[1,2], Chen Ge[1], Jie-su Wang[1], Hai-zhong Guo[1], and Guo-zhen Yang[1,2].

1. *Institute of Physics, Chinese Academy of Sciences, Beijing 100190, China*

2. *Collaborative Innovation Center of Quantum Matter, Beijing 100190, China*

*Correspondence and requests for materials should be addressed to Kui-juan Jin (email: kjjin@iphy.ac.cn).

**Abstract**:

Ferroelectricity has been believed unable to coexist with metallicity since the free carriers can screen the internal coulomb interactions of dipoles. Very recently, one kind of materials called as *ferroelectric metal* was reexamined. Here, we report the coexistence of metallicity and polar distortion in a new candidate for *ferroelectric metal* PbTi$_{1-x}$Nb$_x$O$_3$ via doping engineering. The ferroelectric-like polar distortion in all the doped PbTi$_{1-x}$Nb$_x$O$_3$, with $x$ ranging from 0.04 to 0.12, was confirmed by the piezoresponse force microscopy and the scanning transmission electron microscopy measurements. PbTi$_{1-x}$Nb$_x$O$_3$ films become more conductive with more doping density, and emerge a metallic behavior when $x$ reaches 0.12. Our first principle calculations further revealed that the doped Nb ions in the films can only provide free electrons, but not be able to damage the dipoles in unite cells even with the heaviest doping density of 0.12 due to their little impact on the off-centering of the Ti ions. We believe that these results confirm a feasibility of realizing the coexistence of metallicity and polar distortion for other ferroelectrics in a common way, and motivate the synthesis of some new materials with artificially designed properties even incompatible in nature.

Since Anderson and Blount proposed the statement of "ferroelectric metal" over five decades ago,[1] this type of materials has, mostly, stayed in conceptual or theoretical designing. Such materials should be very hard to obtain because the free carriers screen the long-range coulomb interaction, and therefore weaken the ferroelectric polar distortion. Nevertheless, not long ago, an unambiguous experimental evidence of ferroelectric-like structure transition in metallic material $LiOsO_3$ has reattracted much attention on the metals with non-centrosymmetric structures, which was observed in low temperature below 150 K.[2-8] Yet, they remain challenging to discover.[8,9] So far no room temperature candidate for the coexistence of polar distortion and metallicity has been obtained. Different from the non-centrosymmetric metals, another possible route toward ferroelectric metal is looking for a ferroelectric insulator whose polar distortion survives doping of carriers. Unfortunately, based on the extensive research on the prototypical ferroelectric $BaTiO_3$ with doping engineering, it has been found that the doped electrons has weaken the polar distortion of $BaTiO_3$, or even erased the polar distortion when a critical concentration reached.[10-13] Recently, Raghavan *et al.* pointed out that the ferroelectricity and mobile carriers can coexist in $SmTiO_3/BaTiO_3$ heterostructure by purely electrostatic doping.[14] In addition, $PbTiO_3$ is another common ferroelectric material, but has not been paid enough attention as a doped ferroelectric.[15] Very recently, our theoretical study revealed that $PbTiO_3$ is a wonderful candidate for the doped ferroelectric materials.[16] We have proposed that the lone-pair-driven polar distortion not only persists, but might also be enhanced with electron doping.[16] As reported in Refs. [2-14], doped ferroelectrics and non-centrosymmetric metals are all supposed to be the candidates for ferroelectric metal. However, the ferroelectric-like macroscopic remnant polarization had not been measured out in any of these "ferroelectric metals" before.

Here we present a coexistence of polar distortion and metallicity in Nb doped $PbTiO_3$ at room temperature. In this work, we fabricated the $PbTi_{1-x}Nb_xO_3$ (PTNO) films with various doping concentrations (x = 0.04, 0.06, 0.08, and 0.12) on (001)-oriented $SrTi_{0.993}Nb_{0.007}O_3$ (STNO) conductive single-crystal substrates. The piezoresponse loops were observed by piezoresponse force microscopy (PFM) in films with all above doping densities, in spite of the large leakage currents in the highly doped samples. To confirm the ferroelectricity or the existence of the non-centrosymmetric structure of the PTNO films, a distinct off-centering of the Ti ion was directly characterized by scanning transmission electron microscopy (STEM). Then macroscopic ferroelectric hysteresis loops (*P-E* loops) measurements show that the polarization induced by Ti ion displacements cannot be totally screened by the doped electrons. With increasing x, the temperature dependent behavior of the PTNO films changes from insulating to semiconducting, and further to the metallic. Our first principle calculation further shows that the electron from the doped Nb mainly locate around one of the four nearest Ti ions to Nb and behave itinerantly. Those doped electrons allow the $PbTi_{0.88}Nb_{0.12}O_3$ to present the metal-like behavior, whereas they are not able to effectively impact the covalence bonds of Pb-O and those of Ti-O, so that they cannot damage the off-centering of the Ti ions, neither the individual dipole in each unit cell. We believe that the present work not only realizes a room temperature candidate with the coexistence of polar distortion and metallicity, but also paves a road of doping some ferroelectrics for broadening the class of ferroelectric metals which is with very few candidate so far, but with very unusual properties such as unconventional optical responses,[17,18] magnetoelectricity,[19] and superconductivity,[20,21] .

**Results**

**Polar distortion and ferroelectricity.** In the PTNO films, tetravalent Ti atoms are partly substituted by pentavalent Nb atoms. Due to the similar sizes between $Nb^{5+}$ (r = 0.64 Å) and $Ti^{4+}$ (r = 0.61 Å) ions, substitution of Ti by Nb does not introduce much disorder or lattice distortion in the primary $PbTiO_3$ tetragonal structure.[22] The θ-2θ X-ray diffraction (XRD) and the X-ray absorption spectroscopy (XAS) confirm this $Ti^{4+}$ substitution in our fabricated PTNO films rather than the formation of new compounds (see supporting information Fig. S1).[23] To investigate the ferroelectricity of the PTNO films, ferroelectric domains of them were characterized by PFM. Figs. 1a-d show the out-of-plane PFM phase images, and Figs. 1e-h show the corresponding local piezoresponse phase-voltage loops. The corresponding out-of-plane PFM amplitude images and topography images are given in Fig. S2 a-h (see supporting information). When an alternating (AC) voltage is applied to the surface of a ferroelectric by a conductive tip of PFM, converse piezoelectric effect can induce a resonance of the sample surface, and then the resonance amplitude and phase can be detected by the tip.[24] The change of the resonance phase implies that the polarization changes from one direction to another. Therefore, the areas with different piezoresponse phase can exhibit the different ferroelectric domains. As shown in Fig. 1a, typical strip domains were detected on the surface of the $PbTi_{0.96}Nb_{0.04}O_3$. Little relevance between the phase image and the topography image (as show in Fig. S2a) can eliminate the interference from topographic artifacts. In addition, the local piezoresponse phase-voltage loop in Fig. 1e has a well-defined two states with 180 degree phase difference. We can conclude that the $PbTi_{0.96}Nb_{0.04}O_3$ sample shows a well ferroelectricity in common with the mother compound $PbTiO_3$. For the samples with increasing the concentration of Nb, the $PbTi_{0.94}Nb_{0.06}O_3$ also has an acceptable domain structure in the PFM phase image (Fig. 1b) and has 180 degree phase difference in the local piezoresponse loop (Fig. 1f). However, $PbTi_{0.92}Nb_{0.08}O_3$ and $PbTi_{0.88}Nb_{0.12}O_3$ do not exhibit the same property as those with $x$ less than 0.08. As shown in Fig. 1c, and 1d, the domain structure is amorphous for the $PbTi_{0.92}Nb_{0.08}O_3$ and is even undetectable for the $PbTi_{0.88}Nb_{0.12}O_3$. We believe that the inevitable leakage currents are responsible for the poor PFM images in the heavily doped films. Leakage currents could weaken the electric field applied on the sample, and further affect the piezoresponse signals which were read by PFM. In fact, we have observed the local piezoresponse loops in the $PbTi_{0.92}Nb_{0.08}O_3$ and the $PbTi_{0.88}Nb_{0.12}O_3$ films, even with much noise as shown in fig. 1g and 1h, respectively.

To confirm the existence of polar distortion induced by the Ti/Nb off-centering, , the atomic-scale distinguished STEM imaging was utilized. Figures 2a and 2b represent the aberration-corrected high-angle annular dark-field (HAADF) STEM images of the lightly doped $PbTi_{0.96}Nb_{0.04}O_3$ and the heavily doped $PbTi_{0.88}Nb_{0.12}O_3$, respectively. In the HAADF images, the intensity of atom columns is approximately proportional to $Z^2$ where Z is the atomic number. Therefore, the brightest dots denote the Pb columns and the weaker dots denote the Ti (Nb) columns. From Figs. 2a and 2b, we can observe distinct electric domain structures in both images with 180º and 90º domain walls, which is the characteristic of the pure ferroelectric $PbTiO_3$.[25] The zoom-in STEM images of the two samples are shown in Fig. 2c and 2d, respectively. Obvious displacement of Ti (Nb) ions relative to the center of mass of the four nearest Pb neighbors is marked in these two zoom-in images. For tetragonal ferroelectric $PbTiO_3$, the polarization direction is opposite to the displacement vector of center cations.[25] Based on this, upward polarization and rightward polarization is labeled in Fig. 2c and 2d, respectively. Therefore, we can conclude that the piezoresponse loops in Fig. 1e-h indeed come from the intrinsic polar distortion of the PNTO lattice. More importantly, these results demonstrate that the polar distortion still exists in the heavily doped $PbTi_{0.88}Nb_{0.12}O_3$ samples. It is completely different from the well-studied doped $BaTiO_3$, in which the polar distortion was weakened or vanished by the electron doping.[10-13] In fact, the definition of ferroelectric metals evolved more broad in the previous reports than it firstly comes out decades ago. As reported in Refs. 2-12, doped ferroelectrics and non-centrosymmetric metals are all supposed to be the candidates for ferroelectric metal. However, the ferroelectric-like macroscopic remnant polarization had not been measured out in any of these "ferroelectric metals" before.

In order to prove whether the polar distortion in PTNO can provide a macroscopic remnant polarization or be fully screened by doped electrons, we performed a macroscopic ferroelectricity measurement using a ferroelectric tester. Figure 3a shows the ferroelectric hysteresis loops (*P-E* loops) of Au/PTNO/STNO at ambient temperature. The measured frequency of the *P-E* loops was set to 10 kHz. All of these samples exhibit distinguishable *P-E* loops, in spite of that none of these loops are saturated. We believe the large leakage current is responsible for the unsaturation of *P-E* loops.[26] Anyway, even the heavily doped $PbTi_{0.88}Nb_{0.12}O_3$ has non-zero remnant polarizations when the electric field drops to zero. The *P-E* loops in the Remnant Polarization Mode is shown in Fig. 3b to eliminate the contribution from leakage. Beside we carried out the positive-up negative-down (PUND) measurement to capture the remnant polarization quantitatively. For pulse delay of 1000 ms and pulse width of 1 ms, the twofold remnant polarization (2Pr) of $PbTi_{0.88}Nb_{0.12}O_3$ is 1.54 μC/cm$^2$. This demonstrates that, more than the existence of an internal electric polarization in the heavily doped $PbTi_{0.88}Nb_{0.12}O_3$ films, the polarization charges are not completely screened out by the itinerant electrons. Therefore, it should be found out that whether the dopant Nb atoms provide enough itinerant electrons to transform the conductive behavior of the PTNO films from the insulating into the metallic.

**Temperature-dependent transport behavior.** Pentavalent Nb atoms are regarded as donors which provide free electrons for the $Ti^{4+}$ site substitution in titanate oxides. For examples, an insulating to metallic transition occurs at very small doping levels for SrTi$_{1-x}$Nb$_x$O$_3$ (x<0.0003),[27] however, at much higher doping levels for BaTi$_{1-x}$Nb$_x$O$_3$ (x>0.2).[10] In order to quantitatively evaluate the electrical properties of our doped PTNO films, resistivity versus temperature measurements are implemented. Figure 4 shows the electrical resistivity of PTNO as a function of the temperature. The resistivity becomes slightly smaller with increasing the temperature for PbTi$_{0.96}$Nb$_{0.04}$O$_3$, indicating an insulating behavior. There are, at least, two reasons why the resistivity stays at a large value with slightly doping. Firstly, the electrons liberated from the pentavalent Nb can be balanced by the holes provided by Pb vacancies. Therefore, far from lowering the resistivity, slightly doping of Nb actually increases it.[28] Secondly, the extra electrons provided by Nb are not totally itinerant, instead they are mainly locate around one Ti orbits. This will be further discussed later in the present article based on our first principle calculations. However, for the samples with more dopants, the resistivities of PbTi$_{0.94}$Nb$_{0.06}$O$_3$ and PbTi$_{0.92}$Nb$_{0.08}$O$_3$ decrease remarkably with increasing the temperature. More distinctively, the resistivity of PbTi$_{0.88}$Nb$_{0.12}$O$_3$ goes up with increasing the temperature. Therefore, we can conclude that, with the increase of Nb concentration from 0.04 to 0.12, the temperature-dependent transport behavior transforms from the insulating to the metallic.

**Discussion**

For clarifying the underlying mechanism behind these phenomena, we carried out density functional theory (DFT) calculations (detailed description for the calculation can be found in Experimental Section). For simplicity, the 0.125 instead of 0.12 doping concentration is investigated by constructing a 2×2×2 supercell of PbTiO3 and substituting one Ti by Nb. Generally, doped electrons could localize and form an insulating phase, called as the small polaron[29]. To study the conductive behavior, both delocalized and small polaron configurations need to be modeled. The delocalized solution can be achieved within a standard non-spin-polarized calculation. The small polaron configuration can be investigated by following the strategy proposed by Deskins *et al.*[30] There are six inequivalent positions in the primitive cell of $Pb_8Ti_7NbO_{24}$ where the excess one electron could localize. By taking the polar axis as *c*-axis, we found the most stable solution is that the wave function of the doped electron locates around one of the four Ti atoms nearest to Nb atom in the *ab*-plane as shown in Fig. 5a. Figure 5a also manifests that the spatial distribution of the doped electron wavefunction decays from the localization center. The insulating behavior of our films with the lower doping level than 0.12 can be interpreted as the discretization of energy levels due to the weak wavefunction overlap among the nearest doped electrons. The doped electron is not fully localized to open a band gap in this 0.125 doping case, instead this is a metallic state with some localization as shown in Fig. 5b, which is consistent with our experimental results. In addition, the crystal structure obtained from DFT calculations also show that the polar distortion is not distinctly suppressed by metallicity as demonstrated in Fig. 5c. The $Ti^{4+}/Nb^{5+}$ cations have displacements from the center of Pb8 cage around them, which is in line with the STEM images in Fig. 2. The surviving of the non-centrosymmetric structure of $PbTi_{0.88}Nb_{0.12}O_3$ should mainly because that the *ab-plane* distribution of the doped electron is not able to screen out the dipole in *c* axis.

In summary, we experimentally realized the electron doping in PbTiO$_3$ by fabricating the PTNO films with various Nb concentrations. To verify that the piezoresponse signals stem from the local dipole in the unit cell, HAADF STEM images of the slightly doped PbTi$_{0.96}$Nb$_{0.04}$O$_3$ and the heavily doped PbTi$_{0.88}$Nb$_{0.12}$O$_3$ are used. We found that even the heavily doped PbTi$_{0.88}$Nb$_{0.12}$O$_3$ have a distinguishable domain structure and an off-centering displacement of the Ti$^{4+}$/Nb$^{5+}$ ions. In addition, the existence of a macroscopic remnant polarization measured by ferroelectric tester demonstrates that the polarization charges are not completely screened by the doped electrons in all of these films. The temperature dependent resistivity measurements manifest that, with increasing Nb dopants, the PTNO films underwent a transition from an insulator to a metal. The itinerant electrons provided by Nb atoms present the metal-like behavior, whereas they cannot damage the off-centering of the Ti ions, neither the individual dipole in each unit cell. The presented DFT results has revealed that the doped electrons are not able to effectively impact the orbital hybridization of Pb and O and the hybridization of Ti and O. Thus, PNTO is proved to be with the co-existance of the polar distortion and metallicity at room temperature, as a more promising candidate for ferroelectric metal which is predicted decades ago but has few members up to now. It has been reported that polar metal presents unconventional optical responses, magnetoelectricity, and superconductivity properties.[17-21] We believe that this class of materials has potential applications on functional devices which desire high conductivity coexisting with the polarization field like ferroelectric photovoltaic device. The present results should pave a road of doping some ferroelectrics for broadening this class of ferroelectric metals.

**Method**

**Sample fabrication.** The PbTi$_{1-x}$Nb$_x$O$_3$ (PTNO) films with x = 0.04, 0.06, 0.08, and 0.12 were epitaxial deposited on (001)-oriented SrTi$_{0.993}$Nb$_{0.007}$O$_3$ (STNO) conductive single-crystal substrates by laser molecular beam epitaxy (laser-MBE). During the depositing, we use the commercial PbTi$_{1-x}$Nb$_x$O$_3$ ceramic target materials with x = 0.04, 0.06, 0.08, and 0.12. Therefore the sample stoichiometric proportion should be nearly equal to the target materials. A XeCl 308 nm excimer laser was used with an energy density of 1.5 J cm$^{-2}$ a repetition rate of 2 Hz. During the deposition, the oxygen pressure was approximately 40 mTorr, and the temperature of substrates was 520℃. The thicknesses of the PbTi$_{1-x}$Nb$_x$O$_3$ (PTNO) films are around 200 nm. After the deposition, the samples were *in-situ* annealed for 20 min, and then cooled down to the room temperature.

**Ferroelectric and metallic characterization.** The atom substituted doping and the crystal structures were identified by the x-ray absorption spectroscopy (XAS) and the conventional θ-2θ X-ray diffraction (XRD) method, respectively. The ferroelectric domain structure and local piezoresponse-voltage relationship were characterized by the measurement of piezoresponse force microscopy (PFM) in dual AC resonance tracking (DART) mode.[24] Temperature-dependent resistivity was measured by the combination of a temperature controller (Linkam Scientific Instrument) and a semiconductor parameter analyzer (Keithley 4200), to determine the conductive behavior. For electrical measurements, circular Au electrodes with 100 μm diameter were deposited on the surface of the PTNO/STNO. The ferroelectric hysteresis loops was measured by a ferroelectric test system (Radiant Technologies), to quantitatively evaluate the remanent polarization. Finally, we visualize the polarization of ionic displacement at the atomic scale by using the aberration-corrected high-angle annular dark-field (HAADF) Z-contrast STEM imaging

**Calculation details.** For simplicity, the 0.125 instead of 0.12 doping concentration is investigated by constructing a 2×2×2 supercell of PbTiO$_3$ and substituting one Ti by Nb. Density functional theory (DFT) calculations are performed within the generalized gradient approximation (GGA) on the basis of the projector augmented wave (PAW)[31] method as implemented in the Vienna *ab initio* simulation package (VASP).[32] The exchange and correlation functional parameterized by Perdew-Becke-Erzenhof (PBE)[33] is adopted. The electronic configurations of involved atoms are as follows: Pb ($5d^{10}6s^26p^2$), Nb ($4s^24p^64d^45s^1$), Ti ($2s^22p^63d^24s^2$), and O ($2s^22p^4$). An effective Hubbard term U$_{eff}$ = U–J using Dudarev's approach[34] is included to treat the Ti 3$d$ orbital, and U$_{eff}$ = 3.27 eV generates the most suitable structure compared to the experimental structure. 520 eV energy cutoff of plane wave basis set is taken for all calculations.

**References**

1   Anderson, P. W. & Blount, E. I. Symmetry Considerations on Martensitic Transformations: "Ferroelectric" Metals. *Phys. Rev. Lett.* **14**, 217-219 (1965).

2       Shi, Y. G. *et al.* A ferroelectric-like structural transition in a metal. *Nat. Mater.* **12**, 1024-1027 (2013).

3       Sim, H. & Kim, B. G. First-principles study of octahedral tilting and ferroelectric-like transition in metallicLiOsO3. *Phys. Rev. B* **89**, 201107 (2014).

4       Giovannetti, G. & Capone, M. Dual nature of the ferroelectric and metallic state inLiOsO3. *Phys. Rev. B* **90**, 195113 (2014).

5       Liu, H. M. *et al.* Metallic ferroelectricity induced by anisotropic unscreened Coulomb interaction inLiOsO3. *Phys. Rev. B* **91**, 064104 (2015).

6       Lo Vecchio, I. *et al.* Electronic correlations in the ferroelectric metallic state ofLiOsO3. *Phy. Rev. B* **93**, 161113 (2016).

7       Jin, F. *et al.* Raman phonons in the ferroelectric-like metalLiOsO3. *Phys. Rev. B* **93**, 064303 (2016).

8       Puggioni, D. & Rondinelli, J. M. Designing a robustly metallic noncenstrosymmetric ruthenate oxide with large thermopower anisotropy. *Nat. commun.* **5**, 3432 (2014).

9       Filippetti, A., Fiorentini, V., Ricci, F., Delugas, P. & Iniguez, J. Prediction of a native ferroelectric metal. *Nat. commun.* **7**, 11211 (2016).

10      Liu, L. *et al.* Effects of donor concentration on the electrical properties of Nb-doped BaTiO[sub 3] thin films. *J. Appl. Phys.* **97**, 054102 (2005).

11      Page, K., Kolodiazhnyi, T., Proffen, T., Cheetham, A. K. & Seshadri, R. Local structural origins of the distinct electronic properties of Nb-substituted SrTiO3 and BaTiO3. *Phys. Rev. Lett.* **101**, 205502 (2008).

12      Wang, Y., Liu, X., Burton, J. D., Jaswal, S. S. & Tsymbal, E. Y. Ferroelectric instability under screened Coulomb interactions. *Phys. Rev. Lett.* **109**, 247601 (2012).

13  Iwazaki, Y., Suzuki, T., Mizuno, Y. & Tsuneyuki, S. Doping-induced phase transitions in ferroelectric BaTiO3from first-principles calculations. *Phys. Rev. B* **86**, 214103 (2012).

14  Raghavan, S., Zhang, J. Y., Shoron, O. F. & Stemmer, S. Probing the Metal-Insulator Transition in BaTiO_{3} by Electrostatic Doping. *Phys. Rev. Lett.* **117**, 037602 (2016).

15  Cohen, R. E. ORIGIN OF FERROELECTRICITY IN PEROVSKITE OXIDES. *Nature* **358**, 136-138 (1992).

16  He, X. & Jin, K.-j. Persistence of polar distortion with electron doping in lone-pair driven ferroelectrics. *Phys. Rev. B* **94**, 224107 (2016).

17  Mineev, V. P. & Yoshioka, Y. Optical activity of noncentrosymmetric metals. *Phys. Rev. B* **81**, 094525 (2010).

18  Edelstein, V. M. Features of light reflection off metals with destroyed mirror symmetry. *Phys.Rev. B* **83**, 113109 (2011).

19  Edelstein, V. M. Magnetoelectric effect in dirty superconductors with broken mirror symmetry. *Phys. Rev. B* **72**, 172501 (2005).

20  Bauer, E. *et al.* Unconventional superconducting phase in the weakly correlated noncentrosymmetric Mo3Al2C compound. *Phys. Rev. B* **82**, 064511 (2010).

21  Bauer, E. *et al.* Superconductivity in the complex metallic alloy beta-Al3Mg2. *Phys. Rev. B* **76**, 014528 (2007).

22  Ibrahim, R. C., Horiuchi, T., Shiosaki, T. & Matsushige, K. Highly oriented Nb-doped lead titanate thin films by reactive sputtering: Fabrication and structure analyses. *Jpn. J. Appl. Phys. Part 1 - Regul. Pap. Short Notes Rev. Pap.* **37**, 4539-4543 (1998).

23  Torres-Pardo, A. *et al.* Spectroscopic mapping of local structural distortions in ferroelectric

PbTiO3/SrTiO3superlattices at the unit-cell scale. *Phys. Rev. B* **84**, 220102 (2011).

24   Soergel, E. Piezoresponse force microscopy (PFM). *J. Phys. D: Appl. Phys.* **44**, 464003 (2011).

25   Tang, Y. L. *et al.* Observation of a periodic array of flux-closure quadrants in strained ferroelectric PbTiO3 films. *Science* **348**, 547-551 (2015).

26   Iijima, T., Nafe, H. & Aldinger, F. Ferroelectric properties of Al and Nb doped PbTiO3 thin films prepared by chemical solution deposition process. *Integr. Ferroelectr.* **30**, 9-17 (2000).

27   Frederikse, H. P. R., Thurber, W. R. & Hosler, W. R. Electronic Transport in Strontium Titanate. *Phys. Rev.* **134**, A442-A445 (1964).

28   Ibrahim, R. C., Horiuchi, T., Shiosaki, T. & Matsushige, K. Highly oriented Nb-doped lead titanate thin films by reactive sputtering: Electrical properties. *Jpn. J. Appl. Phys. Part 1 - Regul. Pap. Short Notes Rev. Pap.* **37**, 6060-6064 (1998).

29   Hao, X. F., Wang, Z. M., Schmid, M., Diebold, U. & Franchini, C. Coexistence of trapped and free excess electrons in SrTiO3. *Phys. Rev. B* **91**, 085204 (2015).

30   Deskins, N. A., Rousseau, R. & Dupuis, M. Distribution of Ti3+ Surface Sites in Reduced TiO2. *J. Phys. Chem. C* **115**, 7562-7572 (2011).

31   Kresse, G. & Joubert, D. From ultrasoft pseudopotentials to the projector augmented-wave method. *Phys. Rev. B* **59**, 1758-1775 (1999).

32   Kresse, G. & Furthmuller, J. Efficient iterative schemes for ab initio total-energy calculations using a plane-wave basis set. *Phys. Rev. B* **54**, 11169-11186 (1996).

33   Perdew, J. P., Burke, K. & Ernzerhof, M. Generalized gradient approximation made simple. *Phys. Rev. Lett.* **77**, 3865-3868 (1996).

34   Dudarev, S. L., Botton, G. A., Savrasov, S. Y., Humphreys, C. J. & Sutton, A. P. Electron-energy-

loss spectra and the structural stability of nickel oxide: An LSDA+U study. *Phys. Rev. B* **57**, 1505-1509 (1998).

**Acknowledgements**

The work was supported by the National Key Basic Research Program of China (Nos. 2014CB921001, 2014CB921002 and 2013CB328706), the Key Research Program of Frontier Sciences of the Chinese Academy of Sciences (Grant No. QYZDJ-SSW-SLH020), the Strategic Priority Research Program (B) of the Chinese Academy of Sciences (Grant No. XDB07030200), and the National Natural Science Foundation of China (Grant Nos. 11674385, 11574365, 11474349, 51522212 and 11404380).

**Author contributions**

K-j. Jin supervised the research. J-x. Gu fabricated the samples and conducted the PFM, the ferroelectric tester, and the electrical measurements. Q-h. Zhang and L. Gu performed the TEM measurements. C. Ma performed the first-principle calculations. J-s. Wang, H-z. Guo characterized the crystal structure by the XRD and XAS test. J-x. Gu, K-j. Jin, C. Ge, and G-z Yang wrote and revised the paper. All the authors discussed the results and commented on the manuscript.

**Competing financial interests**

The authors declare no competing financial interests.

**Figure legends**

**Figure 1 | Piezoresponse signal of PbTi$_{1-x}$Nb$_x$O$_3$.** Out-of-plane PFM phase images of the (**a**) PbTi$_{0.96}$Nb$_{0.04}$O$_3$, (**b**) PbTi$_{0.94}$Nb$_{0.06}$O$_3$, (**c**) PbTi$_{0.92}$Nb$_{0.08}$O$_3$, and (**d**) PbTi$_{0.88}$Nb$_{0.12}$O$_3$ films. Corresponding local piezoresponse phase-voltage loops for the (**e**) PbTi$_{0.96}$Nb$_{0.04}$O$_3$, (**f**) PbTi$_{0.94}$Nb$_{0.06}$O$_3$, (**g**) PbTi$_{0.92}$Nb$_{0.08}$O$_3$, and (**h**) PbTi$_{0.88}$Nb$_{0.12}$O$_3$ films.

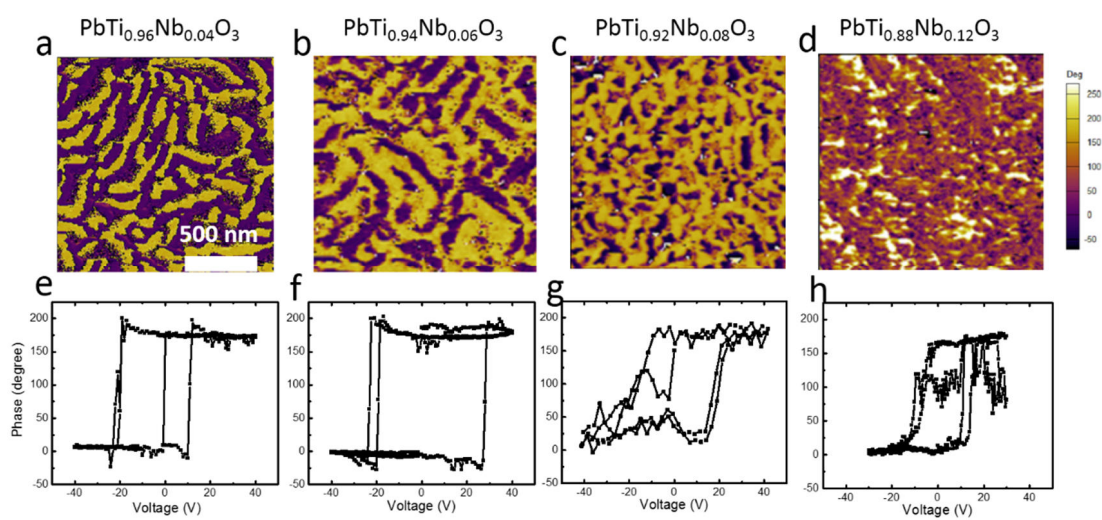

**Figure 2 | Polar-distortion characterization of PbTi$_{0.96}$Nb$_{0.04}$O$_3$ and PbTi$_{0.88}$Nb$_{0.12}$O$_3$.** HAADF-STEM images of (**a**) PbTi$_{0.96}$Nb$_{0.04}$O$_3$ and (**b**) PbTi$_{0.88}$Nb$_{0.12}$O$_3$. Yellow dash lines highlight the domain walls. (**c**) and (**d**) show the zoom-in STEM images corresponding to the areas marked in (**a**) and (**b**), respectively. Red arrows denote the direction of Ti (Nb) displacement with respect to the four nearest Pb. Blue arrows denote the direction of local polarization induced by the ionic displacement.

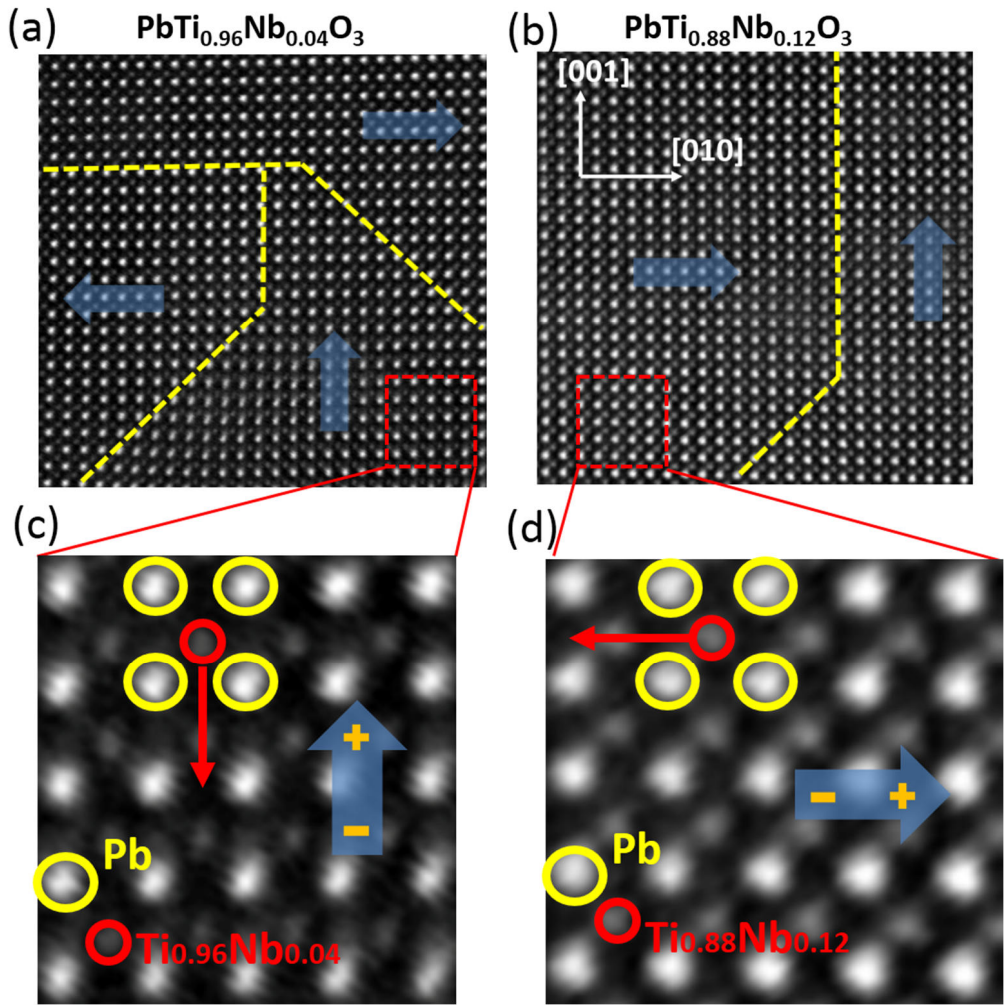

**Figure 3 | Remnant polarization of PbTi$_{1-x}$Nb$_x$O$_3$.** (**a**) Ferroelectric *P-E* loops of the Au/PTNO/STNO heterostructures. (**b**) The *P-E* loops in the Remnant Polarization Mode.

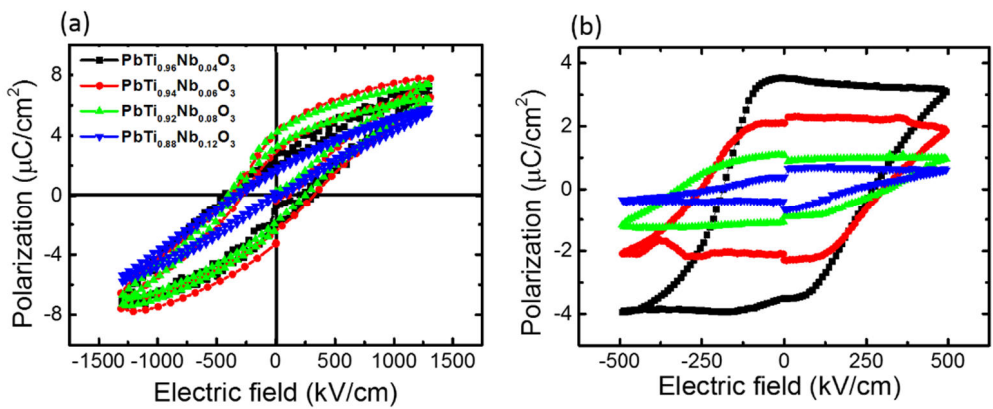

**Figure 4 | Transport behavior.** Temperature dependence of electrical resistivity for PTNO films with Nb concentration varying from 0.04 to 0.12.

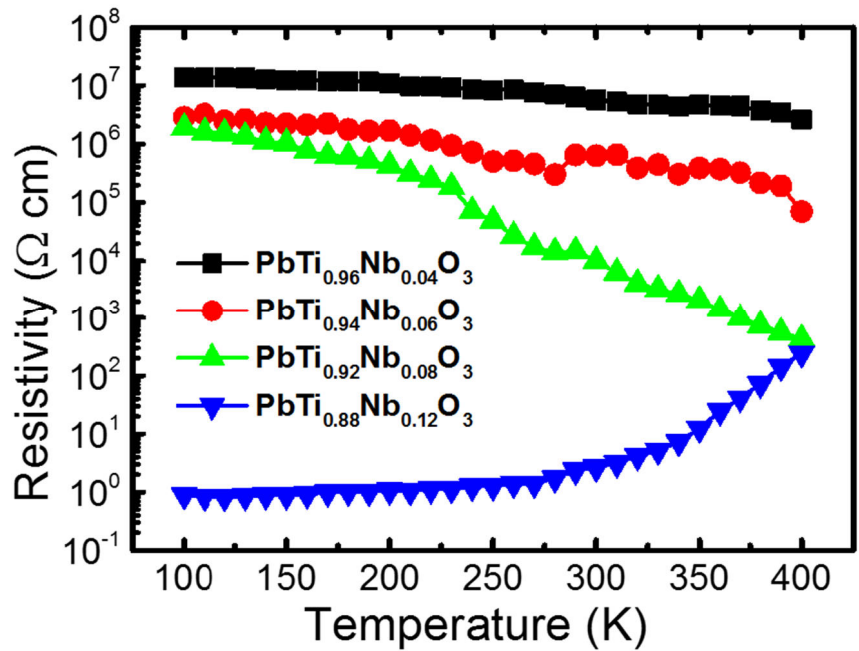

**Figure 5 | Theoretical calculations.** a) The isosurface (isovalue 0.04 eA$^{-3}$) of the doped electronic density for the most stable solution in the Pb$_8$Ti$_7$NbO$_{24}$ primitive cell. b) The total density of states. The yellow area denotes the doped electronic density of states. c) Side view of the Pb$_8$Ti$_7$NbO$_{24}$ primitive cell obtained from DFT.

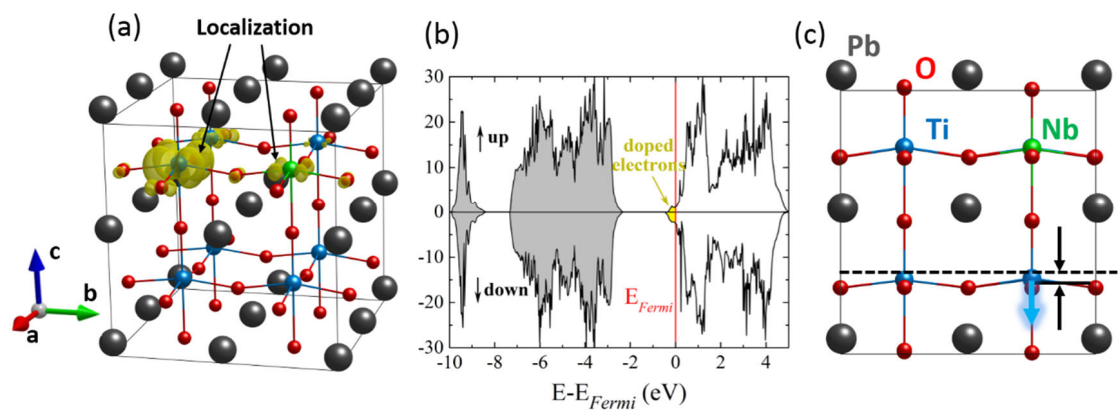